\documentclass{PoS}

\usepackage{epsfig}

\PoS{PoS(LAT2005)097}

\title{Excited meson spectroscopy with chirally improved 
fermions\footnote{For the Bern-Graz-Regensburg (BGR) Collaboration.}  }

\ShortTitle{Excited meson spectroscopy with CI fermions }

\author{\speaker{T.~Burch}, C.~Hagen,
        D.~Hierl, and A.~Sch\"afer\\

	Institut f\"ur Theoretische Physik\\
        Universit\"at Regensburg\\
	D-93040 Regensburg, Germany.\\

        E-mail: \email{tommy.burch@physik.uni-r.de}\\
       }

\author{Christof Gattringer, L.~Ya.~Glozman\footnote{Supported by Fonds zur
    F\"orderung der Wissenschaftlichen Forschung in \"Osterreich, 
    project P16823-N08.} , and C.~B.~Lang\\
	Institut f\"ur Physik, FB Theoretische Physik\\
        Karl-Franzens-Universit\"at Graz\\
	A-8010 Graz, Austria.\\
       }

\abstract{
We present excited meson masses from quenched calculations using chirally 
improved (CI) quarks at pion masses down to 350 MeV. 
The salient features of our analysis are the use of a matrix of correlators 
from various source and sink operators and a basis which includes quark 
sources with different spatial widths, thereby improving overlap with states 
exhibiting radial excitations.
}

\FullConference{XXIIIrd International Symposium on Lattice Field Theory\\
		 25-30 July 2005\\
		 Trinity College, Dublin, Ireland}

\begin{document}

\section{Introduction}

The spectroscopy of excited hadrons still poses a significant problem to 
the lattice QCD community. 
Since a Euclidean-time hadron correlator is composed of a sum of 
contributions from the different mass eigenstates, each of which being 
exponentially suppressed by a factor of the mass, disentangling 
the higher mass states from those of the dominant ground state (let alone 
the other excited masses) can be a daunting task.

In the following, we present a method (see also \cite{BuGaGl04}) for 
handling just this ``disentanglement'' of masses and we show our latest 
results for the meson sector. 
In another contribution to this conference \cite{CHtalk}, 
we present the results for baryons.

\section{Method}

Our method is based upon the variational approach developed by 
Michael \cite{Mi85} and later refined by L\"uscher and Wolff \cite{LuWo90}. 
We use a number of source and sink interpolators to create a matrix of 
correlators which we then diagonalize.

The basic structure of our meson interpolators is 
\begin{equation}
O \; = \; \overline{u}_{n,w} \; \Gamma \; d_{n,w} \; ,
\end{equation}
where we use $\Gamma =$ $\gamma_5$ and $\gamma_4 \gamma_5$, 
$\gamma_i$ and $\gamma_4 \gamma_i$, $\gamma_5 \gamma_i$, 
$\gamma_i \gamma_j$, and $1$ for pseudoscalar (PS), vector (V), 
pseudovector (PV), 
tensor (T), and scalar (SC) mesons, respectively. The subscripts $n$ and $w$ 
refer to narrow and wide quark sources. These are created using 
gauge-covariant Jacobi smearing to approximate Gaussian distributions of 
widths 0.27 and 0.41 fm. By opening this new degree of freedom to the mesons, 
we hope to improve overlap with states exhibiting radial excitations 
(a difference in sign between the narrow and wide source contributions will 
create a radial node in the corresponding quark wavefunction). 
We use degenerate $u$ and $d$ quark masses. 
For strange mesons, we replace one of the light quarks by the strange quark 
(the physical strange quark mass is determined via the PS kaon).

For our non-strange PS and V mesons we then have up to 6 different 
interpolators ($nw$ = $wn$ for degenerate quark masses) for our basis, 
giving up to a $6 \times 6$ correlator matrix (this becomes $8 \times 8$ for 
the strange mesons). 
For the PV, T, and SC mesons, we have a basis of up to 3 interpolators 
(4 for the strange mesons).

After creating our correlator matrix, $C(t)$, we must then solve the 
generalized eigenvalue problem:
\begin{equation}
C(t) \, \vec{v}^{(k)} \; \; = \; \; \lambda^{(k)}(t) \; C(t_0) \, 
\vec{v}^{(k)} \; ,
\label{generalized}
\end{equation}
where the $\lambda^{(k)}(t)$ are the eigenvalues corresponding to the 
eigenvectors $\vec{v}^{(k)}$. 
We note in passing that we arrive at almost identical eigenvalues, up to a 
constant factor, with a normal eigenvalue problem, i.e., by omitting the 
matrix $C(t_0)$ on the r.h.s. (the eigenvectors, however, 
are a different matter). 
The eigenvalues provide the time dependence, and hence the mass, of the 
separate mass eigenstates: 
\begin{equation}
\lambda^{(k)}(t) \; \propto \; e^{-t \, M_k} \,[ \, 1 + 
{\cal O}(e^{-t \, \Delta M_k}) \,] \; .
\label{eigenvaluedecay}
\end{equation}
The eigenvectors provide the corresponding wavefunctions (within the highly 
limited basis).
The ${\cal O}(e^{-t \, \Delta M_k})$ terms in Eq.\ (\ref{eigenvaluedecay}) 
can be reduced by changing, even by further limiting, the starting basis. 
We observe such effects in our results and, in the end, see the best 
effective mass plateaus in the excited-state eigenvalues by choosing 
$3 \times 3$ correlator matrices throughout
($n \Gamma n$, $n \Gamma w$, $w \Gamma w$; 
the only exceptions are for the PS and V mesons on our coarse lattice, where 
we use 4 basis states: 
$n \Gamma n$, $n \gamma_4 \Gamma n$, $n \gamma_4 \Gamma w$, $w \gamma_4 \Gamma w$).

We have results from two sets of quenched configurations, created using the 
L\"uscher-Weisz gauge action. The lattice spacings are determined via the 
Sommer parameter \cite{GaHoSc02}: $a=0.148$ fm for the $16^3 \times 32$ and 
$a=0.119$ fm for the $20^3 \times 32$ (both spatial volumes are $\approx 
2.4$ fm). 
The quark propagators are created using the chirally improved (CI) action 
\cite{CIref}.

\begin{figure}[t]
\begin{center}
\includegraphics[width=11cm,clip,angle=270]{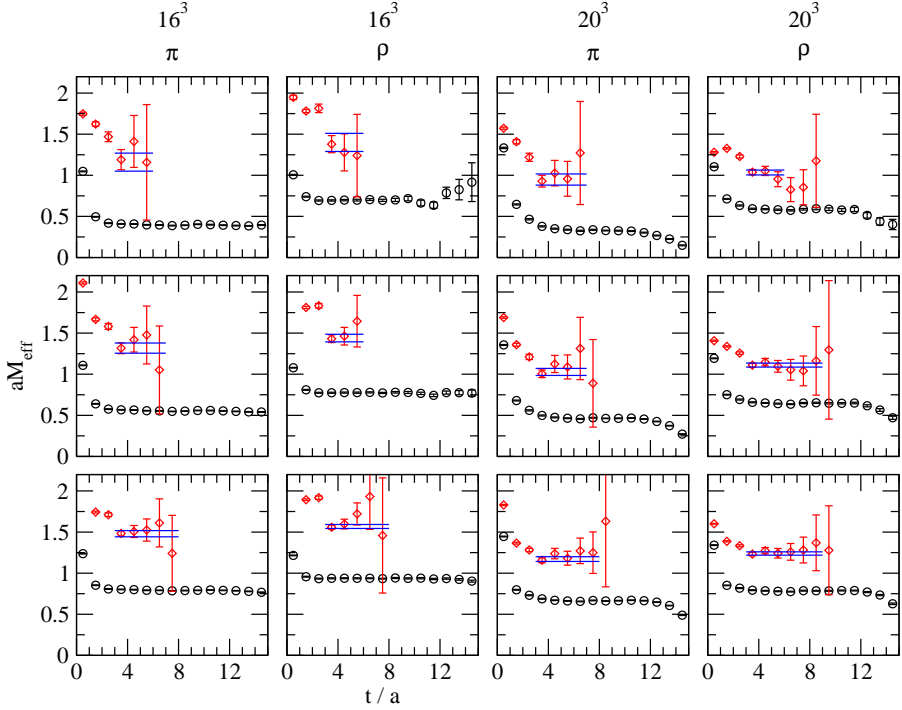}
\caption{ Effective mass plots for pseudoscalar and vector mesons from our 
coarse ($16^3\times 32$, with $am_q=0.05,0.1,0.2$ from top to bottom) and fine 
($20^3 \times 32$, with $am_q=0.04,0.08,0.16$) lattices. 
Both ground and excited states are shown, along with the $M\pm\sigma_M^{}$ 
results (horizontal lines) from fits of the eigenvalues
in the corresponding time intervals. 
\label{eff_mass_ps_vt}}
\end{center}
\end{figure}

\section{Results}

In Fig.\ \ref{eff_mass_ps_vt}, we show some of our effective mass plots for 
the PS and V mesons. 
Results are shown for both the first and second eigenvalues, corresponding 
to the ground and first-excited states. 
From the apparent plateaus, we choose appropriate time intervals over which we 
fit the eigenvalues with an exponential and extract the mass.

\begin{figure}[t]
\begin{center}
\includegraphics[width=7cm,height=7cm]{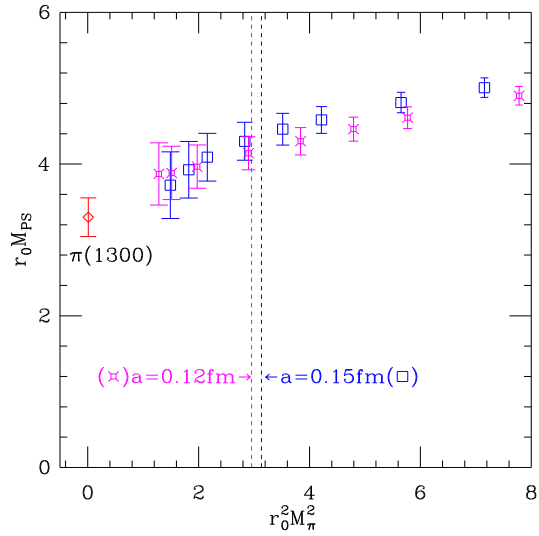}
\caption{
Excited-state pseudoscalar masses vs $M_\pi^2$ for both 
lattice spacings. The quark masses are degenerate. 
All quantities are in units of the Sommer parameter $r_0$. 
The diamond represents the experimental point and the vertical lines mark 
the values of $r_0^2 M_\pi^2$ corresponding to the physical strange quark 
mass.
\label{ps_ud_vs_ps2}}
\end{center}
\end{figure}

Repeating this excercise for many quark masses, we arrive at plots like that 
of Fig.\ \ref{ps_ud_vs_ps2}, where the excited-state PS mass is 
displayed as a function of the ground-state PS mass squared ($\propto m_q$). 
Although the statistical errors are substantial, the masses appear to be 
extrapolating to a value consistent with the physical one. 
Also, the results from the two lattice spacings appear to be consistent with 
each other, suggesting small discretization effects for this state.

\begin{figure}[t]
\includegraphics[width=7cm]{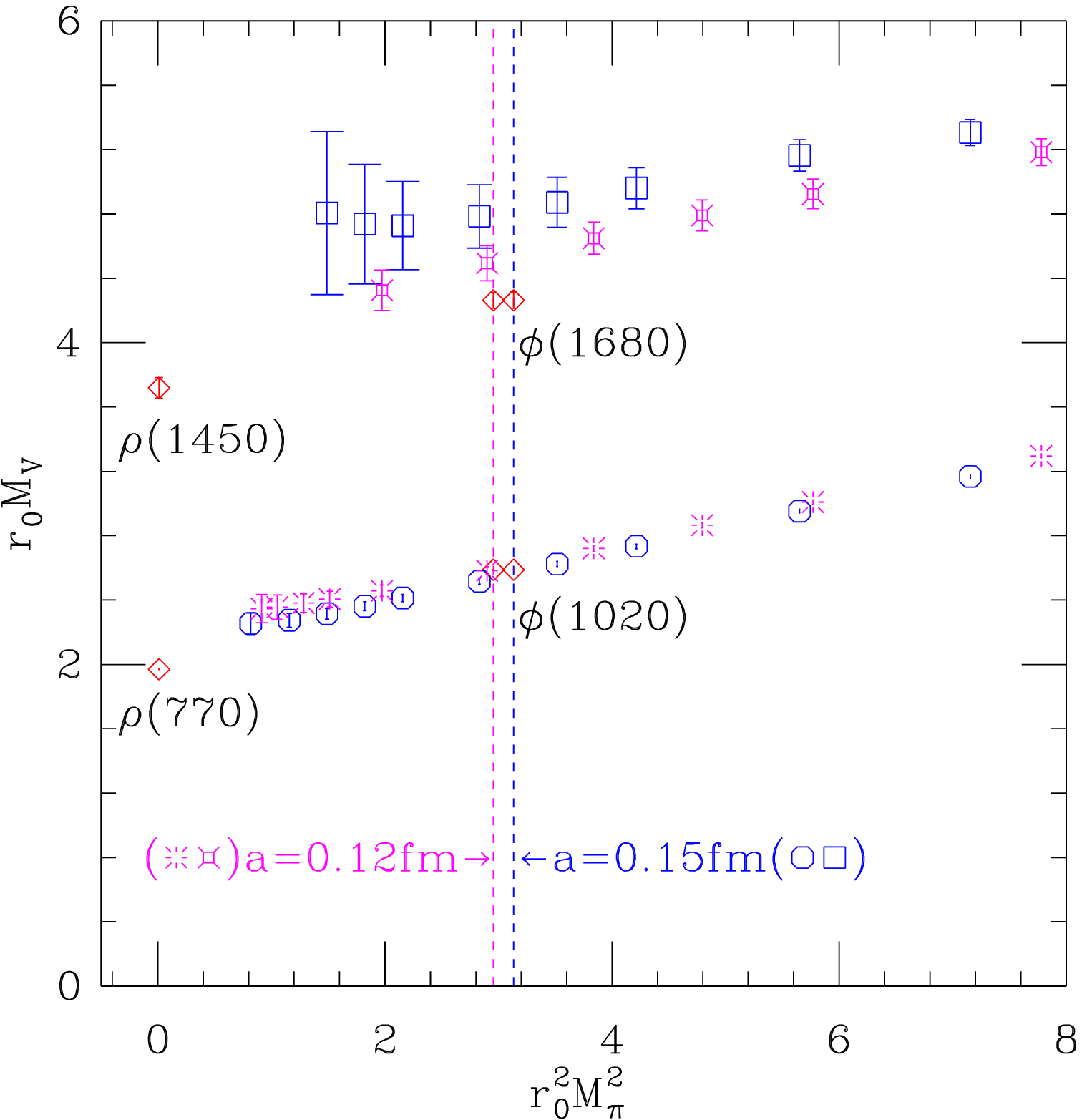}
\hspace{\fill}
\includegraphics[width=7cm]{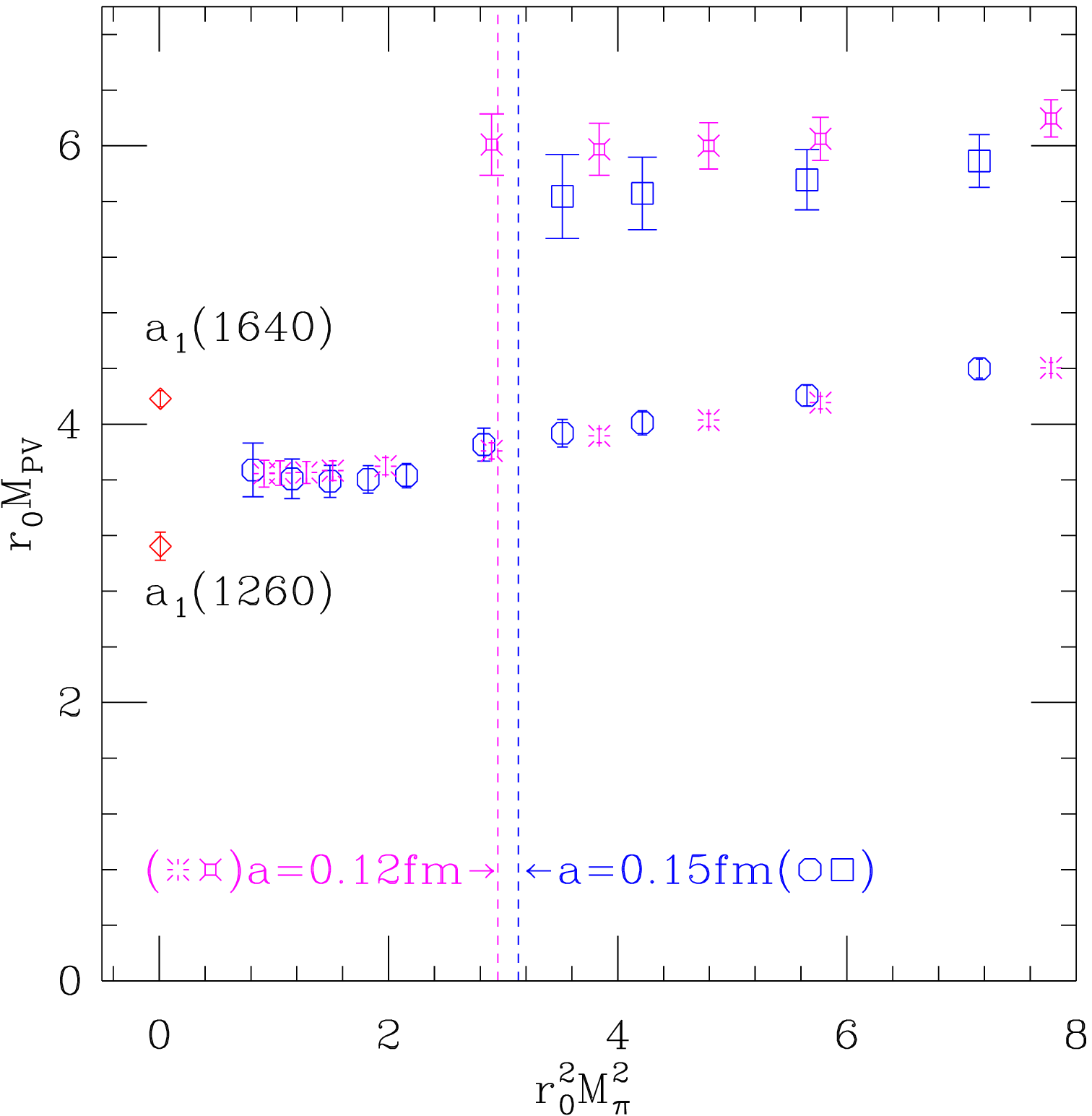}\\

\includegraphics[width=7cm]{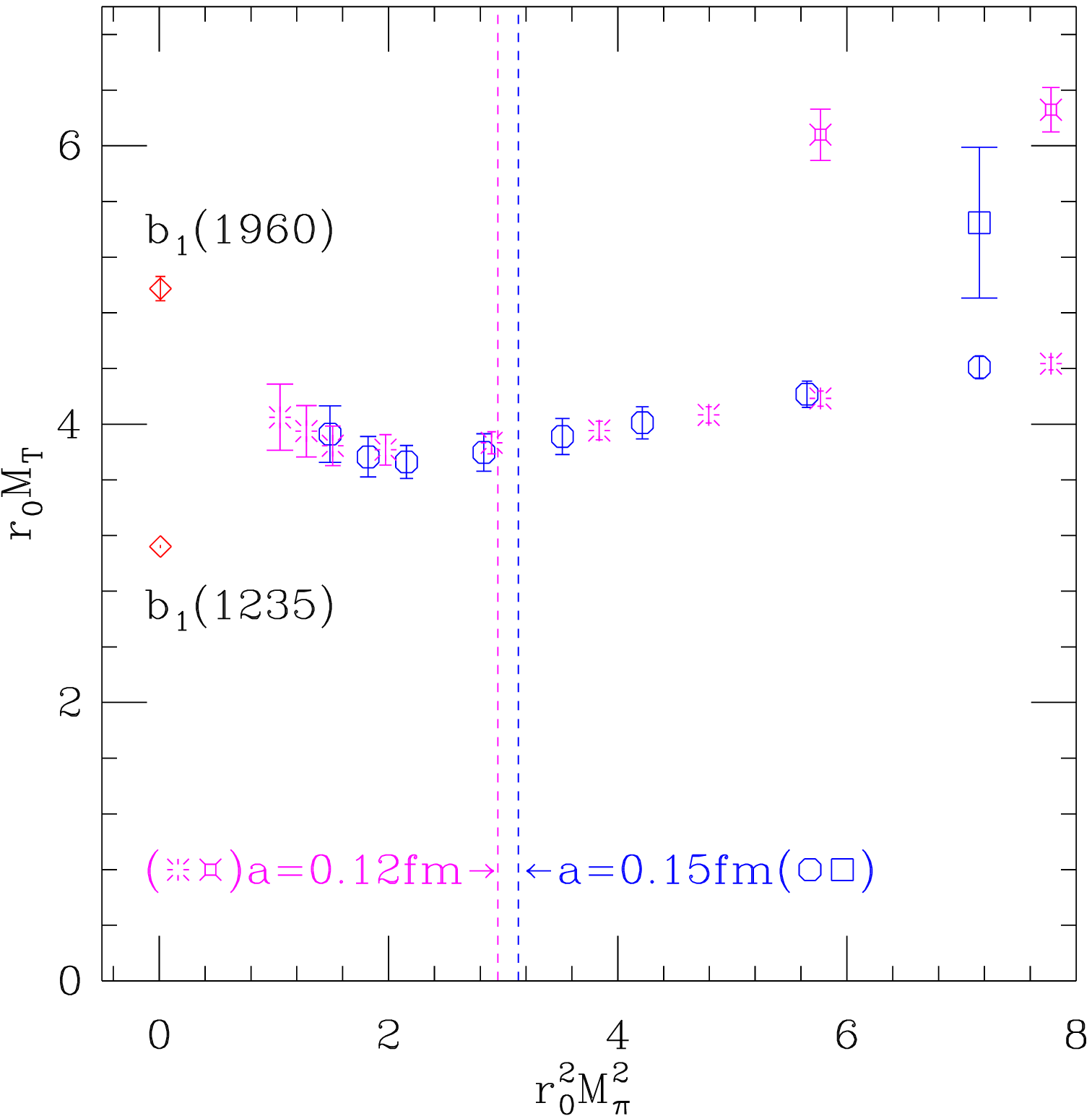}
\hspace{\fill}
\includegraphics[width=7cm]{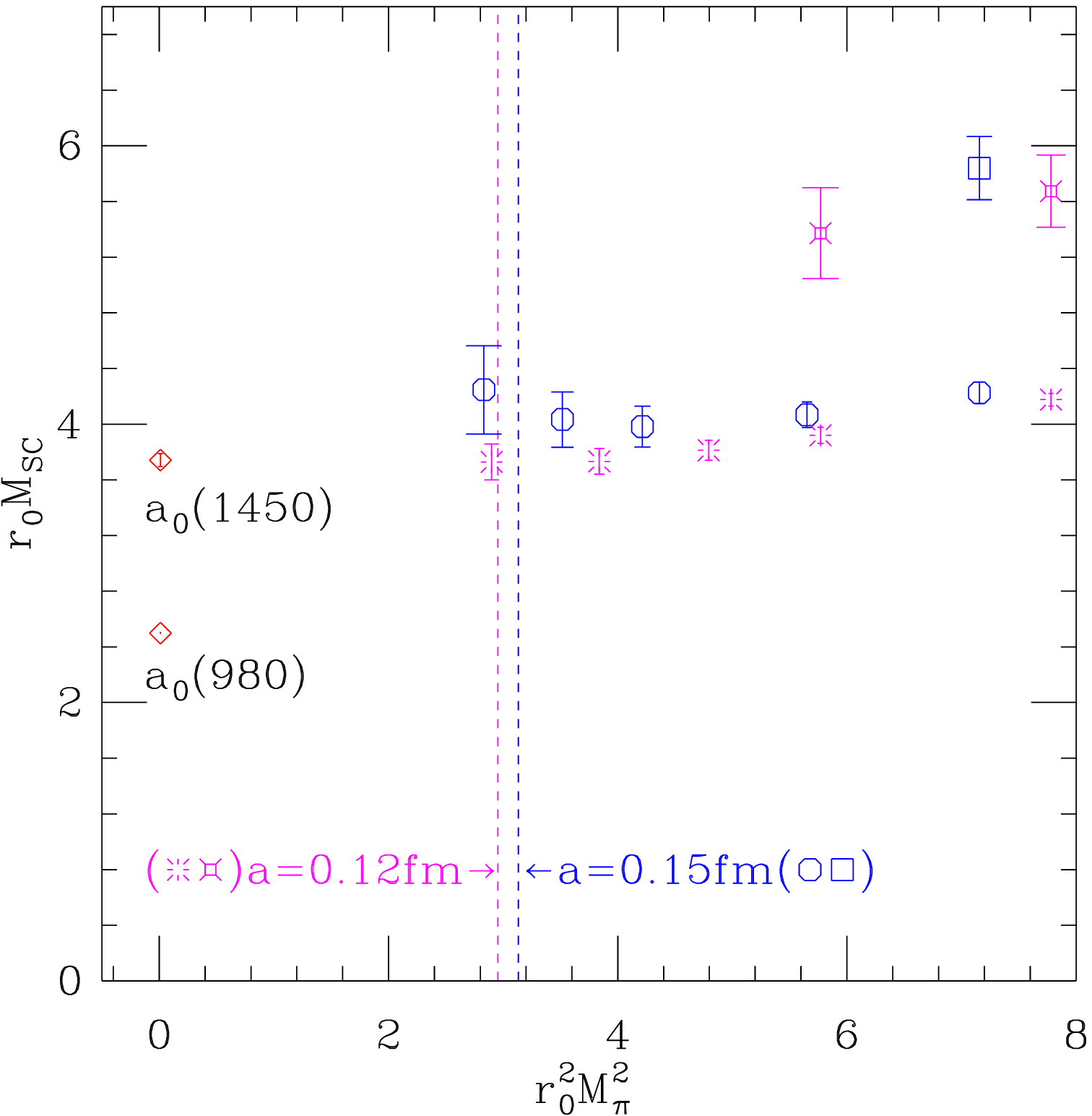}
\caption{
Ground- and excited-state meson masses vs $M_\pi^2$ for both 
lattice spacings. Vector, pseudovector, tensor, and scalar mesons appear. 
The quark masses are degenerate. 
All quantities are in units of the Sommer parameter $r_0$. 
The diamonds represent the experimental points and the vertical lines mark 
the values of $r_0^2 M_\pi^2$ corresponding to the physical strange quark 
mass.
\label{vpts_ud_vs_ps2}}
\end{figure}

In Fig.\ \ref{vpts_ud_vs_ps2}, similar plots are shown for the V, PV, T, and 
SC states. 
For the V mesons, the results are roughly consistent with the experimental 
values, although it is apparent that discretization effects are larger for 
these excited states. 
The other states are more problematic. 
The excited PV is obviously higher than one might naively expect and 
results for the other excited states are virtually non-existent (few reliable 
effective mass plateaus could be found). 
We plan to improve the picture for such states by including more 
``realistic'' interpolators (e.g., p-wave sources). 
For the SC meson at small quark masses, ghost states dominate the
correlation function (which then becomes negative \cite{bardeen}), and 
we therefore plot only the results at large quark masses.

\begin{figure}[t]
\includegraphics[width=7cm]{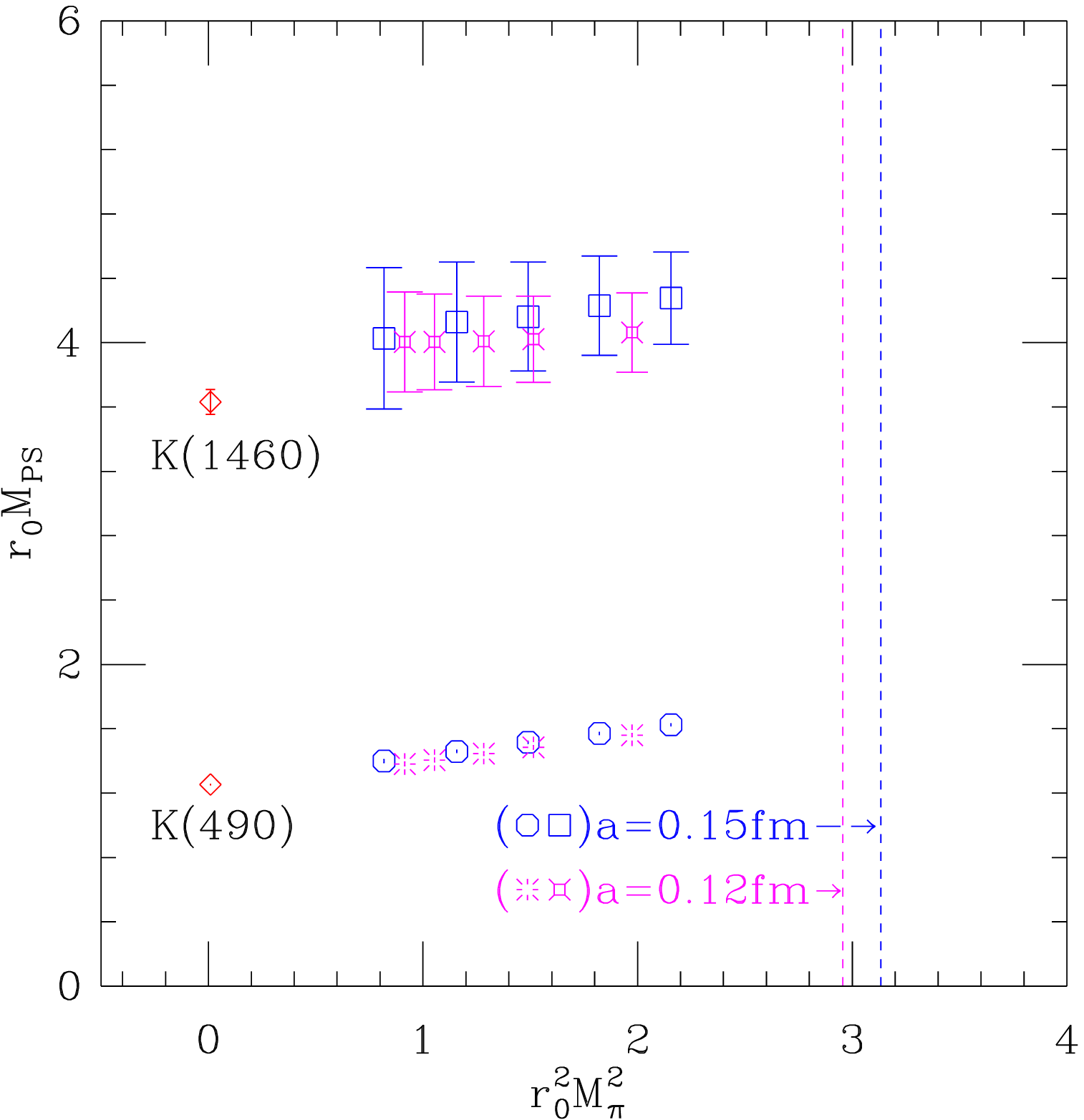}
\hspace{\fill}
\includegraphics[width=7cm]{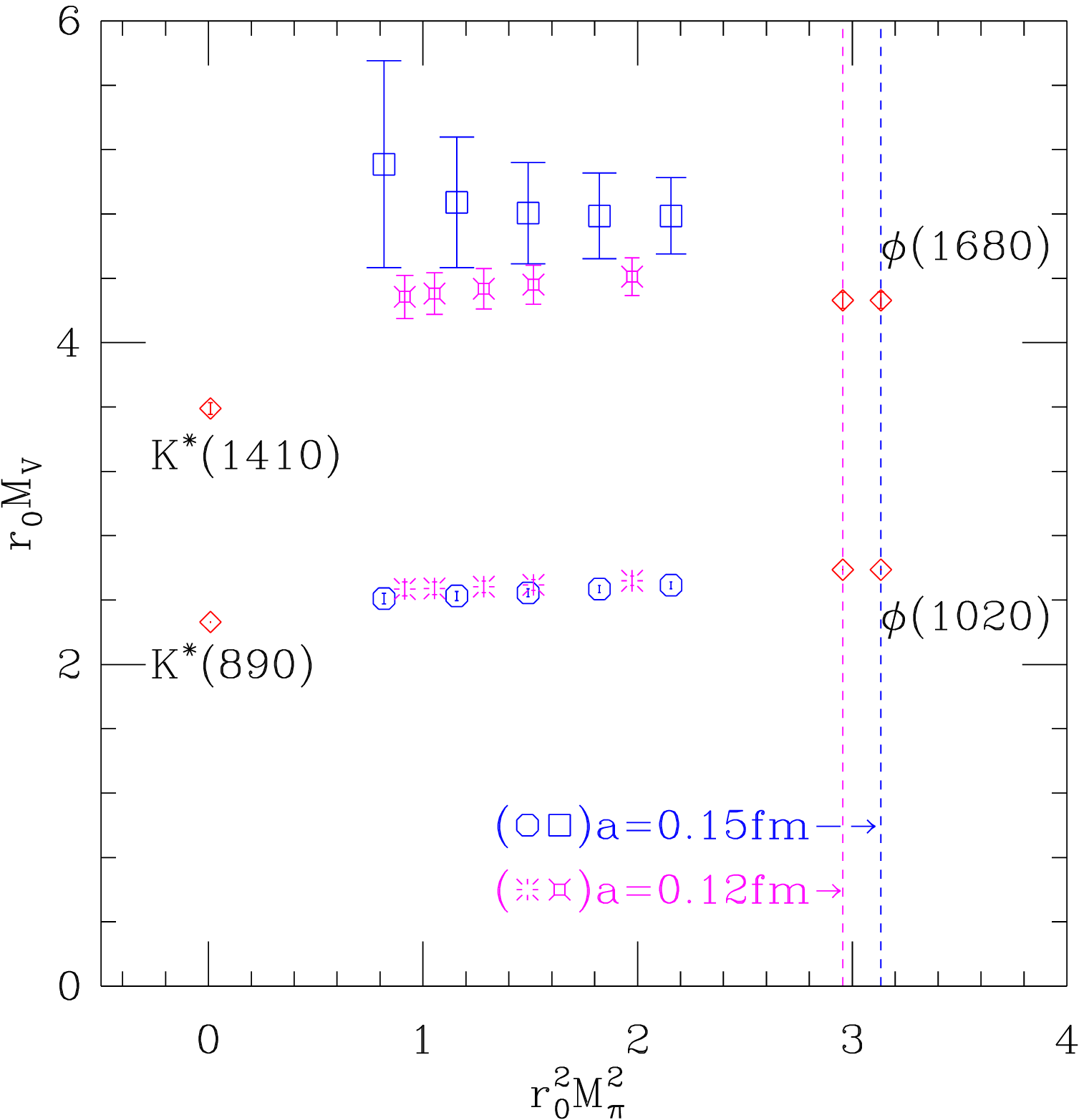}
\caption{ Ground- and excited-state meson masses vs $M_\pi^2$ for both 
lattice spacings. Both pseudoscalar and vector mesons are shown. 
One quark mass is fixed to the physical strange quark mass. 
All quantities are in units of the Sommer parameter $r_0$. 
The diamonds represent the experimental points and the vertical lines mark 
the values of $r_0^2 M_\pi^2$ corresponding to the physical strange quark 
mass.
\label{ps_vt_us_vs_ps2}}
\end{figure}

Figure \ref{ps_vt_us_vs_ps2} shows the same plots for the strange PS and V 
mesons. 
Again, the excited PS results are consistent with the experimental value and 
display small discretization effects. 
The excited V mesons, however, appear to remain high and also suffer more 
from discretization effects.

When looking at the eigenvectors in the $3 \times 3$ smearing basis alone,
invariably we find the same sign for all  components in the ground state but
sign differences for the excited state.  In other words, within this limited
basis, the excited state has a radial  node in its wavefunction.  Including
different spin structures ($\Gamma$ and $\gamma_4 \Gamma$) in the PS or the V
basis, the interpretation becomes more ambiguous.

Of course, all of these results are from quenched lattices with relatively 
small volumes ($\approx 2.4$ fm). So we have some systematic effects which are 
rather difficult to quantify without repeating our analysis on larger, 
dynamical configurations. It is quite possible that the small volume may be 
the reason the excited V and PV mesons remain higher than the experimental 
values.

\end{document}